%% file: make_astro.tex
\documentclass{mpe_report}

\usepackage{psfig,graphicx,epsfig}
\usepackage{color}
\usepackage{amsmath,amssymb,epic,eepic,array}

\begin{document}

\pagenumbering{arabic}
\setcounter{page}{100}

\renewcommand{\FirstPageOfPaper }{100}\renewcommand{\LastPageOfPaper }{103}\include{./mpe_report_stappers}        \clearpage

\end{document}

%% file: mpe_report_stappers.tex

\title{LOFAR: A Powerful Tool for Pulsar Studies}
\author{B. W. Stappers\inst{1}, A. G. J. van Leeuwen\inst{2}, M. Kramer\inst{3}, D. Stinebring\inst{4}, J. Hessels\inst{5}}
\institute{Stichting ASTRON, Postbus 2, 7990 AA Dwingeloo, The Netherlands
\and Department of Physics and Astronomy, University of British Columbia, 6224 Agricultural Road, Vancouver, B.C. V6T 1Z1, Canada.
\and University of Manchester, Jodrell Bank Observatory, Macclesfield, Cheshire SK11 9DL, UK.
\and Department of Physics and Astronomy, Oberlin College, Oberlin, OH 44074
\and Department of Physics, McGill University, Montreal, QC H3A 2T8, Canada.}

\authorrunning{Stappers et al.}
\maketitle

\begin{abstract}
The LOw Frequency ARray, LOFAR, will have the sensitivity, bandwidth,
frequency range and processing power to revolutionise low-frequency
pulsar studies. We present results of simulations that indicate that a LOFAR survey
will find approximately 1500 new pulsars. These new pulsars will give
us a much better understanding of the low end of the luminosity
function and thus allow for a much more precise estimate of the true
total local pulsar population. The survey will also be very sensitive
to the ultra-steep spectrum pulsars, RRATs, and the pulsed radio emission
from objects like Geminga and AXPs. We will also show that by enabling
us to observe single pulses from hundreds of pulsars, including many 
millisecond pulsars, LOFAR opens up new possibilities for the study
of emission physics.
\end{abstract}

\section{Introduction}

LOFAR will provide  a low frequency radio telescope concentrated in the
Netherlands with extensions into other European countries.  It
consists of a core, an extended array and a long baseline component
which will have maximum baselines of 2~km, 100~km and $\sim$1000~km
respectively. It will operate in the frequency range of 30 -- 240~MHz
which will be split into two bands, the low band (optimised for 30--80
MHz) and the high band (optimised for 115-240 MHz). Bandwidths of
either 100 or 80~MHz will be digitised with 12-bit resolution and it
will be possible to select a total of 32 MHz of bandwidth to be
processed. There will be a total of 7700 low-band antennae and 7700
high-band tiles where each tile is made up of a grid of 16
antennae. The antennae and tiles will be divided up into a total of 77
stations. In the core there will be 32 stations and the remaining 45
will be arranged in a logarithmic spiral-like distribution outside of
the core. All the antennae in a station will be combined to form a
so-called station beam and it will be possible to exchange bandwidth
for beams. For example one could decide to have 8 station beams (each
with an independent look direction within the field of view of the
antenna or tile) each of 4 MHz of bandwidth. This multibeaming
capability allows for extremely wide fields of view to be monitored
and/or it allows for multiple experiments to be carried out
simultaneously. Estimates of the LOFAR sensitivity for pulsar
observations using the core are given in Table \ref{params} and more
information about LOFAR in general can be found at {\tt www.lofar.org}

\begin{table}
      \caption{LOFAR Sensitivity for Pulsar observations$^{\rm a}$}
         \label{params}
      \[
         \begin{array}{p{0.5\linewidth}rr}
            \hline
            \noalign{\smallskip}
             Frequency   &  Sensitivity & Beam Size \\
             (MHz)   &  (mJy)  &  \\
            \noalign{\smallskip}
            \hline
            \noalign{\smallskip}
             30  & 0.48 & 21^{'} \\
             75  & 0.33 & 8.3^{'} \\
             120  & 0.017 & 5.2^{'} \\
             200  & 0.015 & 3.1^{'} \\
            \noalign{\smallskip}
             \hline
         \end{array}
      \]
\begin{list}{}{}
\item[$^{\rm a}$] Integration times of 1~h, 2 polarisations, 32 MHz of bandwidth, 10\% duty cycle and only using the LOFAR core.
\end{list}
   \end{table} 
  
\begin{figure*}
\centerline{\psfig{file=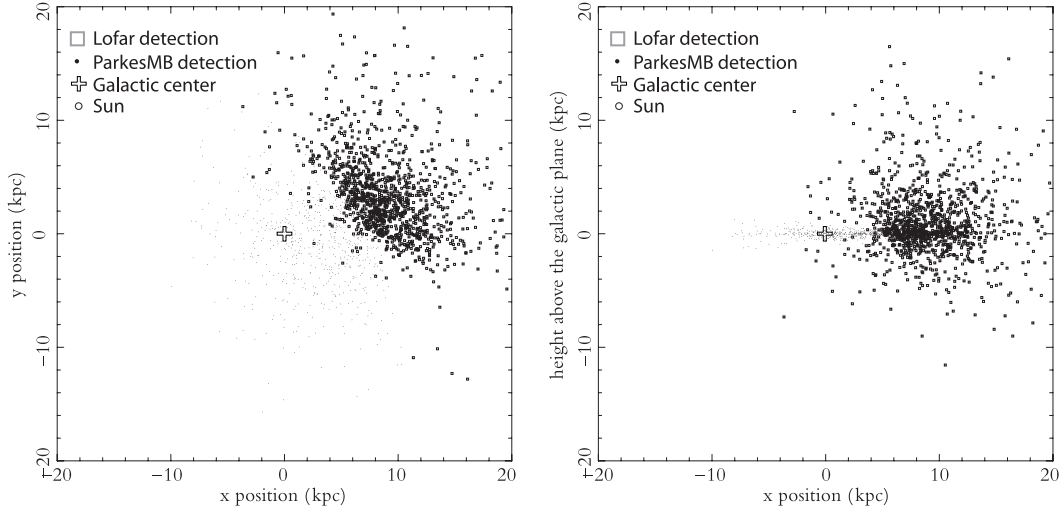,width=14cm,clip=} }
\caption{ Simulated detections for the LOFAR and Parkes Multibeam
     surveys, for 1-hour LOFAR pointings. Left) projected on the
     Galactic plane. Right) projected on the plane through the
     Galactic centre and sun, perpendicular to the disk.
\label{x-y-z}}
\end{figure*}
                 
\section{Pulsar Survey:}

Pulsars are steep spectrum objects whose pulsed flux density usually
peaks in the 100-200 MHz range (e.g. Malofeev et al
1994\nocite{mgj+94}). As LOFAR will have unprecedented sensitivity in
exactly this frequency range this makes it an excellent instrument for
carrying out an all-sky pulsar survey. How many pulsars LOFAR will
discover depends on the low end of the pulsar luminosity
function. There are indications that the luminosity function turns
over in the range 0.3-1 mJy kpc$^2$ \cite{lml+98} however it is
likely that surveys have so far been incomplete already at the 10 mJy
kpc$^2$ level. LOFAR will be able to measure the low-end of the pulsar
luminosity function significantly better than any previous survey
allowing, for the first time, a much more precise estimate of the true
total local pulsar population.

The apparent detection of a few neutron stars including Anomalous
X-ray Pulsars and Magnetars only at frequencies near or below 100 MHz
(e.g. Malofeev et al 2005, Kuzmin \& Losovsky 1997, Malofeev \& Malov
1997r, Shitov \& Pugachev 1997)\nocite{mm97,mmt+05,kl97,sp97b}, the
existence of pulsars like B0943+10, which have flux density spectra
with a spectral index steeper than -3.0 \cite{dr94}, and millisecond
pulsars which have steep spectra which do not show a turn-over even at
frequencies as low as 30 MHz (e.g.~\cite{nbf+95}), suggest that a
large number of pulsars is detectable only at low
frequencies. Moreover, the radio beam of pulsars is known to broaden
at low frequencies, increasing the illuminated sky and hence the
detection probability at Earth. Indeed, the non-detection of Geminga
at cm-radio wavelengths, a prominent rotation-powered pulsar visible
at optical, X-ray and gamma-ray wavelengths, serves as a good example
for a population of neutron stars that may be detectable only with
LOFAR.  With its sensitivity to these very different types of neutron
stars, a LOFAR Galactic pulsar survey will be instrumental in
providing a complete picture of neutron-star radio emission.

We have carried out sophisticated simulations using realistic
population and scattering models to determine the optimum observing
configuration, observing strategy and frequency for a pulsar survey
with LOFAR (Figure \ref{x-y-z} and van Leeuwen \& Stappers 2006). If
we extrapolate the known low-luminosity tail, we find that a LOFAR
all-sky pulsar survey (the first survey of the Northern hemisphere in
10 years) will find around 1500 new pulsars, almost doubling the total
number of pulsars known.  The survey provides a complete local census
of radio-emitting neutron stars, such as radio pulsars, AXPs and
previous "radio-quiet" pulsars.  Using the derived population
properties (such as the pulsar period distribution) one can study the
birth properties of neutrons stars, core collapse physics, the
velocities and spatial distribution of pulsars, and the physics of
neutron stars in general. We note also that the long pointings
possible, because of the wide field of view, in a LOFAR pulsar survey,
makes it particularly sensitive to the recently discovered classes of
infrequently emitting neutron stars like RRATs \cite{mll+06} and
pulsars that are on for 10\% or less of the time \cite{klo+06}.

\section{A survey for extragalactic pulsars.}

Spiral and irregular galaxies will host young, bright, Crab-like
pulsars, and due to its sensitivity, LOFAR can be the first telescope
to find such pulsars besides those in the Magellanic Clouds. If
observed face-on and located away from the Galactic disk, the scatter
broadening to an external galaxy will be relatively low and thus a
LOFAR survey will have excellent sensitivity for pulsars of all spin
periods (van Leeuwen \& Stappers 2006). For a relatively close galaxy
like M33, LOFAR could detect all pulsars more luminous than ~50Jy
kpc$^2$. Our Galaxy hosts 10 pulsars which are brighter than
that. There are at least 20 Galaxies for which LOFAR will have good
sensitivity to their pulsar population. Complementary to this normal
pulsar emission, some pulsars show ultra-bright 'giant pulses' that
could be visible in even more remote galaxies (e.g.~\cite{mc03}).

A survey for extragalactic pulsars would allow us to investigate if
the bright end of the pulsar distribution in other galaxies differs
from our galaxy, and how that ties into galaxy type and star formation
history. Such pulsars can also help the understanding of the missing
baryon problem, the history of massive star formation in these
galaxies and also if sufficient numbers can be found they can be used
to probe the turbulent inter-galactic medium.

\begin{figure}
\centerline{\psfig{file=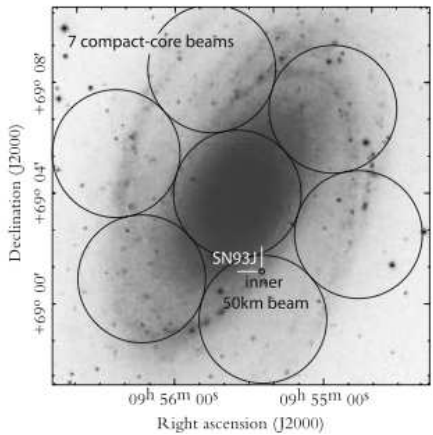,width=6cm,clip=}}
\caption{ Coherently formed beams from the compact core and the inner
  12-km of LOFAR, projected on the core of M81, the second best
  candidate galaxy for an extragalactic survey and also the home of
  supernova 1993J.
\label{M33}}
\end{figure}

\section{Pulsar emission physics}

The sensitivity and frequency range of LOFAR open up the low-frequency
window of pulsar emission to exciting new studies. It is precisely in
the LOFAR frequency range where some of the most interesting changes
in pulsar radio emission can be observed, including the very
significant broadening of the pulse profile \cite{cor78}, changes in
the form of the pulse profile (\cite{kis+98,ran83}), a turn-over in the
flux density spectrum, and it is also the frequency range where
propagation effects in the pulsar magnetosphere may be expected to be
largest (e.g.~\cite{cr79,pet01}). Studying the latter, in particular
with simultaneous multi-frequency observations of single pulses, will
reveal interesting characteristics of the pulsar magnetosphere, such
as particle densities and birefringence properties, that will
ultimately lead to a better understanding of the workings of pulsars.

Our calculations show that in the LOFAR high band single pulses can
potentially be detected with reasonable signal to noise from up to
one-third of pulsars while in the low band it is about one-quarter of
pulsars. This is a very large increase on what has previously been
possible, and when combined with LOFAR's ability to track sources,
allows for the rich study of the emission physics of radio
pulsars. Furthermore LOFAR will also be able to detect single pulses
from millisecond pulsars (MSPs), something which has so far been done
for less than a handful of sources (e.g.~\cite{es03b}), and will allow
us for the first time to compare the mode changing, nulling and
drifting subpulse properties of MSPs with their younger, higher
magnetic field, slower spinning brethren.

Single pulse studies at low frequencies are particularly interesting,
because microstructure ( see Popov et al. 2002\nocite{pbc+02} and
references therein) and subpulses tend to be stronger there. Indeed,
we would expect the density imbalances and plasma dynamics to be the
most noticeable at low radio frequencies. The duration and separation
timescales of microstructure and subpulse emission are similar to
those predicted by theoretical models of pulsar emission physics.  As
the exact timescales are depending on the physics the detailed study
of this type of emission can better differentiate between models of
pulsar emission. While the high time resolution requirement of these
observations and the existence of interstellar scattering will limit
the number of sources that can be studied, the remaining number of
many tens of sources for which studies can be made, makes this a
particular exciting aspects of LOFAR radio studies.

Combining simultaneous multi-frequency observations of radio pulsars
using multiple telescopes has proved very successful at revealing much
about the processes discussed above (e.g.~\cite{khk+01,kkg+03}). 
Unfortunately such studies at low frequencies have been hampered by 
the fact the majority of low-frequency facilities available for this work are 
transit instruments and so only short simultaneous observations are
possible. Using the frequency range and tracking ability of LOFAR in
observations simultaneous with high frequency instruments will further
open up this extremely rich source of information on the way in which
pulsars work.

\begin{figure}
\centerline{\psfig{file=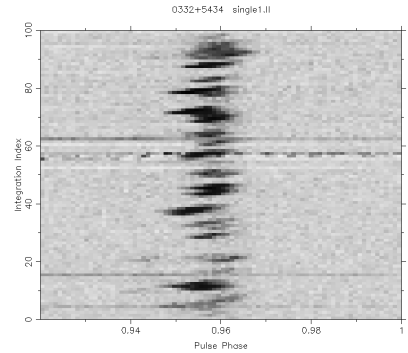,width=6cm,clip=}}

\caption{A stack of single pulses of PSR B0329+54 obtained with the
  Westerbork Synthesis Radio Telescope at a frequency of 143 MHz with
  a bandwidth of 2.5 MHz. The typical signal-to-noise ratio for each
  individual pulse in these data is 30.  LOFAR will have a factor of
  at least 30 greater sensitivity, up to 12 times the bandwidth and
  greater robustness to interference thereby greatly increasing the
  number of systems for which such studies can be undertaken.}

\label{image}
\end{figure}

\section{Interstellar medium studies}

Pulsars are excellent probes of the ionized component of the
interstellar medium through scintillation \cite{ric01}, dispersion
measure, and Faraday rotation studies (e.g.~\cite{hmql03}).
Scintillation studies have been revolutionized in the last five years
by the discovery of faint halos of scattered light extending outward
10--50 times the width of the core of the scattered image
\cite{smc+01}.  This, in turn, gives a wide-angle view of the
scattering medium with milliarcsecond resolution, and the illuminated
patch scans rapidly across the scattering material because of the high
pulsar space velocity. LOFAR will be an outstanding instrument with
which to study this phenomenon, particularly for strong, nearby
pulsars. The dynamic nature of these phenomena also fits well with
LOFAR's excellent monitoring capabilities.  In a sense, this
scintillation imaging will allow us to monitor the range of
interstellar conditions encountered along a particular sight line.
This monitoring will not be particularly time consuming, but it will
add greatly to the more static view of the ISM afforded through other
techniques.

LOFAR will also make important contributions to traditional dispersion
measure, rotation measure, and scattering measure determinations.  By almost
doubling the number of known pulsars, the proposed survey will add a dense
grid of new sight lines through the Galaxy to those that already exist. The
dispersion and scattering measures of this new sample will improve our global
model of the distribution of the ionized ISM and its degree of clumpiness.
The new rotation measures will place important constraints on the overall
magnetic field structure of the Milky Way, which is still not well
characterized. At LOFAR frequencies it is possible to also measure the very
small rotation measures of the nearby population of pulsars providing an
unprecedented tool for studying the local magnetic field structure.

\section{Other possibilities}

LOFAR will be very useful for monitoring and timing. While it will not
perform high precision timing, at the top of the high-band
sufficiently accurate timing will be possible to enable, for example,
frequent observations of a large sample of glitching pulsars. This
would allow for the better parameterisation of glitches and the
possibility of triggering rapidly after a glitch to allow follow-up at
other wavelengths. Similarly timing of a large numbers of pulsars will
also be valuable for follow up with GLAST where accurate ephemerides
are required to fold the long data sets. With regular monitoring it
will also be possible to study and potentially catch transitions in the
"more off than on" pulsars. The radio-sky monitor will also find new
transient pulsar sources, such as RRATs, and allow rapid follow-up and
via the transient buffer board a look back in time at the event
itself.

\section{Conclusions}

LOFAR will reopen, with wider bandwidths, large frequency range and
unprecedented sensitivity, the low-frequency window for pulsar
emission studies. The wide fields of view and sensitivity will also
allow for an efficient and sensitive survey of the whole Northern sky
enabling the discovery of hundreds of new pulsars including a number
of exotic objects and systems. A survey of local group galaxies will also
likely detect the first truly extragalactic pulsars. Pulsars observed with 
LOFAR will provide excellent probes of the interstellar, and potentially 
intergalactic, medium and will be especially useful for gaining a better 
understanding of the local interstellar medium. Furthermore the 
exciting possibilities of multibeaming and radio-sky monitoring open up 
new avenues of pulsar and radio emitting neutron star research.



%% file: make_astro.bbl
\begin{thebibliography}{23}
\expandafter\ifx\csname natexlab\endcsname\relax\def\natexlab#1{#1}\fi

\bibitem[{Cheng \& Ruderman 1979 }]{cr79}
Cheng, A.~F. \& Ruderman, M. 1979, ApJ, 229, 348

\bibitem[{(Cordes 1978)}]{cor78}
Cordes, J.~M. 1978, ApJ, 222, 1006

\bibitem[{(Deshpande \& Radhakrishnan 1994)}]{dr94}
Deshpande, A.~A. \& Radhakrishnan, V. 1994, Journal of Astrophysics and
  Astronomy, 15, 329

\bibitem[{Edwards} \& {Stappers} 2003]{es03b}
{Edwards}, R.~T. \& {Stappers}, B.~W. 2003, A\&A, 407, 273

\bibitem[{Han} {et~al.} 2003]
{hmql03}
{Han}, J.~L., {Manchester}, R.~N., {Qiao}, G.~J., \& {Lyne}, A.~G. 2003, in ASP 
Conf. Ser. 302: Radio Pulsars, ed. M.~{Bailes}, D.~J. {Nice}, \& S.~E. {Thorsett}, 253

\bibitem[{Karastergiou} {et~al.} 2001]
{khk+01} 
{Karastergiou}, A., {von Hoensbroech}, A., {Kramer}, M., {et~al.} 2001, A\&A,
  379, 270

\bibitem[{Kramer} {et~al.} 2003]
{kkg+03}
{Kramer}, M., {Karastergiou}, A., {Gupta}, Y., {et~al.} 2003, A\&A, 407, 655

\bibitem[({Kramer} {et~al.} 2006)]
{klo+06}
{Kramer}, M., {Lyne}, A.~G., {O'Brien}, J.~T., {Jordan}, C.~A., \& {Lorimer},
  D.~R. 2006, Science, 312, 549

\bibitem[{Kuzmin} {et~al.} 1998]
{kis+98} 
Kuzmin, A.~D., Izvekova, V.~A., Shitov, Y.~P., {et~al.} 1998, A\&AS, 127, 255

\bibitem[{Kuzmin \& Losovskii(1997)}]{kl97}
Kuzmin, A.~D. \& Losovskii, B.~Y. 1997, Astronomy Letters, 23, 283

\bibitem[({Lyne} {et~al.} 1998)]
{lml+98}
Lyne, A.~G., Manchester, R.~N., Lorimer, D.~R., {et~al.} 1998, MNRAS, 295, 743

\bibitem[({Malofeev} {et~al.} 1994)]
{mgj+94}
Malofeev, V.~M., Gil, J.~A., Jessner, A., {et~al.} 1994, A\&A, 285, 201

\bibitem[{Malofeev \& Malov(1997)}]{mm97}
Malofeev, V.~M. \& Malov, O.~I. 1997, Nature, 389, 697

\bibitem[{Malofeev} {et~al.}(2005)]
{mmt+05}
{Malofeev}, V.~M., {Malov}, O.~I., {Teplykh}, D.~A., {Tyul'Bashev}, S.~A., \&
  {Tyul'Basheva}, G.~E. 2005, Astronomy Reports, 49, 242
  
\bibitem[{McLaughlin \& Cordes 2003}]{mc03}
McLaughlin, M.~A. \& Cordes, J.~M. 2003, ApJ, 596, 982

\bibitem[({McLaughlin} {et~al.} 2006)]
{mll+06}
{McLaughlin}, M.~A., {Lyne}, A.~G., {Lorimer}, D.~R., {et~al.} 2006, Nature,
  439, 817

\bibitem[{Navarro} {et~al.} 1995]
{nbf+95}
Navarro, J., de~Bruyn, G., Frail, D., Kulkarni, S.~R., \& Lyne, A.~G. 1995,
  ApJ, 455, L55

\bibitem[{Petrova} 2001]{pet01}
{Petrova}, S.~A. 2001, A\&A, 378, 883

\bibitem[{Popov} {et~al.}(2002)]
{pbc+02}
{Popov}, M.~V., {Bartel}, N., {Cannon}, W.~H., {et~al.} 2002, A\&A, 396, 171

\bibitem[{Rankin 1983}]{ran83}
Rankin, J.~M. 1983, ApJ, 274, 333

\bibitem[({Rickett} 2001)]{ric01}
{Rickett}, B. 2001, 278, 129

\bibitem[{Shitov \& Pugachev (1997)}]{sp97b}
Shitov, Y.~P. \& Pugachev, V.~D. 1997, New Astr., 3, 101

\bibitem[({Stinebring} {et~al.} 2001)]
{smc+01}
{Stinebring}, D.~R., {McLaughlin}, M.~A., {Cordes}, J.~M., {et~al.} 2001, ApJ,
  549, L97

\end{thebibliography}
